\documentclass[twoside,english]{iopart}
\usepackage{geometry}
\usepackage{color}
\usepackage{textcomp}
\usepackage{amsbsy}
\usepackage{graphicx}
\usepackage{esint}

\makeatletter
\usepackage{iopams}
\usepackage{setstack}

\usepackage[numbers, sort&compress]{natbib}

\def\newblock{\hskip .11em plus .33em minus .07em}

\newcommand{\eqref}[1]{(\ref{#1})}

\makeatother

\usepackage{babel}
\begin{document}

\title{Dark Field Differential Dynamic Microscopy enables the accurate characterization
of the roto-translational dynamics of bacteria and colloidal clusters}

\title[Dark Field DDM characterization of roto-translational diffusion]{}

\author{{\normalsize{}Roberto Cerbino$^{1}$, Davide Piotti$^{1}$, Marco
Buscaglia$^{1}$, and Fabio Giavazzi$^{1}$}}

\address{1 Dipartimento di Biotecnologie Mediche e Medicina Traslazionale,
Universit� degli Studi di Milano, via F.lli Cervi 93, 20090 Segrate,
Italy}

\ead{roberto.cerbino@unimi.it, fabio.giavazzi@unimi.it}
\begin{abstract}
Micro- and nanoscale objects with anisotropic shape are key components
of a variety of biological systems and inert complex materials, and
represent fundamental building blocks of novel self-assembly strategies.
The time scale of their thermal motion is set by their translational
and rotational diffusion coefficients, whose measurement may become
difficult for relatively large particles with small optical contrast.
Here we show that Dark Field Differential Dynamic Microscopy is the
ideal tool for probing the roto-translational Brownian motion of shape
anisotropic particles. We demonstrate our approach by successful application
to aqueous dispersions of non-motile bacteria and of colloidal aggregates
of spherical particles.
\end{abstract}

\maketitle

\section{Introduction}

Understanding and quantifying Brownian processes is relevant for soft
condensed matter scientists as well as for a wider audience that ranges
from biologists to economists \cite{Haw:2002sf,Bian:2016vn}. As far
as colloidal particles are of interest, the erratic nature of their
Brownian motion is reflected in the well-known fractal appearance
of their trajectories as well as in the irregular change of their
orientation in time \cite{Doi:2011gf,Doi:2013mz}. In the past, rotational
Brownian motion has received considerably less attention than its
translational counterpart, in part because characterizing the rotational
Brownian motion is more challenging. Most of the characterization
makes use of optical methods such as Video Particle Tracking (VPT)
, Dynamic Light Scattering (DLS), and Fluorescence Correlation Spectroscopy
(FCS) \cite{Piazza:1990fr,Berne:2000ye,borsali2008soft,TRACK2008}.
In particular, Depolarized Dynamic Light Scattering (D-DLS) has been
shown to be a powerful tool to assess the roto-translational dynamics
of an ensemble of anisotropic (by shape and/or optically) particles
by analyzing the fluctuations in the depolarized scattered light intensity
\cite{Berne:2000ye,Eimer1990,Lehner2000,Piazza:1990fr,Balog2014}.
Very recently, it was shown that D-DLS experiments, usually requiring
a custom laser-based optical setup, can be performed successfully
with an optical microscope \cite{Giavazzi:2016_jpcm}. This approach,
termed polarized-Differential Dynamic Microscopy (p-DDM), builds on
Differential Dynamic Microscopy (DDM) that extracts scattering information
from the quantitative analysis of time-lapse microscope movies \cite{Cerbino:2008if}.
While DDM in its original implementation probes the translational
diffusion coefficient of a colloidal suspension observed in bright
field microscopy, p-DDM gives also access to the rotational diffusion
of anisotropic colloidal particles, provided that the sample is observed
between suitably oriented polarizers. Compared to DLS and D-DLS, DDM
and p-DDM offer a simpler implementation, more robust performances,
higher flexibility, and a better rejection of both stray and multiply-scattered
light \cite{Cerbino:2009ta,Giavazzi:2014sj,cerbino_cicuta2017}.

Both D-DLS and p-DDM work well with spherical particles made of a
markedly birefringent material and with non-spherical (optically homogeneous)
particles provided that both the shape anisotropy and the optical
contrast with the dispersion medium are large enough. These requirements
are not always met and, as a result, the rotational dynamics of important
classes of particles cannot be probed with D-DLS and p-DDM. In particular,
particles that are quasi-index matched with the dispersion liquid
or sparse aggregates exhibit a depolarized component of the scattered
light that is insufficiently large for computing reliably the temporal
autocorrelation function of the scattered field (p-DDM) or intensity
(D-DLS). A quick and reliable tool for the quantitative characterization
of the roto-translational dynamics of such samples is not yet available
but it would be useful not only for the great potential of anisotropic
particles in the self assembly arena \cite{SOLOMON_natmat} but also
because many systems of biological interest such as bacteria and prokaryotic
cells cannot be easily studied with D-DLS \cite{RGD1979}.

An alternative to D-DLS for the characterization of the rotational
dynamics of shape anisotropic particles is available, at least for
particles that are not too small. In fact, for an anisotropic particle
whose longest dimension $L$ is comparable or larger than the inverse
scattering wave-vector ($L\gtrsim150$ $nm$, typically), the intensity
of the light scattered at a large angle depends upon the orientation
of the particle itself and from the study of the fluctuations in the
scattered light intensity one can measure the so called dynamic form
factor \cite{Pecora:1968bh}, which encodes information on the rotational
dynamics of the particles, as well as on other internal degrees of
freedom, if present \cite{Berne:2000ye}. 

Here we show that Dark Field Differential Dynamic Microscopy (d-DDM),
a recently proposed method for the characterization of the translational
diffusion of colloidal particles \cite{Bayles:2016_darkDDM}, enables
also the effective quantification of the intensity fluctuation associated
with the rotational dynamics. Our key observation is that, in a conventional
dark field microscope, the image of a small particle is mainly formed
by light scattered at relatively large angles, while forward scattered
light is not collected. This implies that the intensity associated
with the dark field image of an anisotropic particle depends on its
orientation. Accordingly, if the particle rotates, for example because
of thermal motion, the intensity of the collected light fluctuates,
leading to an image that ``blinks'' over time. This makes d-DDM
a valuable tool also for the quantitative characterization of the
roto-translational dynamics of colloidal particles, especially in
regimes where p-DDM does not work, namely the case of micron-sized
particles closely index-matched with the solvent. To demonstrate this
approach we first present measurements on a dispersion of non-motile
rod-shaped bacteria. The determination of the rotational and translational
diffusion coefficients allows to determine the relevant dimensions
of the bacterial particles. In a second set of experiments, d-DDM
is applied to a suspension of spherical colloidal particles. In this
case d-DDM is demonstrated to be extremely sensitive to the presence
of even a small fraction of aggregates of particles, whose anisotropic
nature leads to a significant rotational signal. 

\section{Materials and methods}

\subsection{Sample preparation}

In order to assess the potential of d-DDM in characterizing the roto-translational
dynamics of anisotropic colloidal particles, we have employed two
different kind of samples that are difficult to study with other tools
such as p-DDM and/or D-DLS. In particular, we studied aqueous dispersions
of non-motile bacteria and of aggregates of spherical nanoparticles,
both presenting shape anisotropy and a moderate refractive index difference
with the dispersion medium (water). 

\subsubsection*{Non-motile bacteria}

One of the most frequently used E. coli strain for routine biological
cloning applications is the DH5$\alpha$ strain, which are non flagellated
bacteria. They are almost non-motile, less fragile to handle and easy
to grow, which makes them the perfect anisotropic particles for our
purposes. We grew single colonies from frozen stocks on Luria broth
(LB) agar plates at $37$ \textdegree C overnight. A single colony
was transferred from a plate to $20$ ml of liquid LB and incubated
overnight ($16$ h) at $37{^\circ}C$ while shaken (for aeration)
at 200 rpm. The next step was to transfer cells from LB to a minimal
medium with no exogenous nutrients in order not to have bacterial
reproduction and to minimize the growth rate. To this aim, the medium
underwent a washing process consisting of a 2 min centrifugation (6000
rpm and 2000 g), the expulsion of supernatant and the resuspension
of bacterial pellet in 1 ml of PBS (phosphate-buffered saline, a water-based
salt solution containing sodium hydrogen phosphate and sodium chloride).
This washing procedure was repeated 3 times (2 centrifugations). After
this, the system was sucked into a $0.3\times1\times20$ mm capillary
and the capillary was fixed to a glass slide with vaseline petroleum
jelly, a non-toxic glue. This operation was carefully performed in
order not to let air get inside the capillary that could lead to unwanted
sample drifts. It was also important to carefully avoid liquid residuals
between the capillary and the glass slide, which would have caused
a dynamic drying front expanding during the microscope acquisition
and affecting the DDM experiments.

To avoid bacteria sedimentation, which impact on the dynamics, the
density mismatch between bacteria and the PBS physiological medium
was compensated by adding Percoll$\circledR$ to the dispersion. Percoll$\circledR$
is routinely used for density gradient centrifugation of cells, viruses,
and sub-cellular particles. It consists of colloidal silica particles
of $15-30$ nm diameter prepared at $23$\% w/w in water. The silica
particles are coated with polyvinylpyrrolidone (PVP), which makes
them completely non-toxic and ideal for use with biological materials.
PVP is randomly bound to the silica particles as a monomolecular layer.
The size of these particles is so small that the intensity of the
scattered light is negligible when compared to the intensity of the
light scattered by bacteria, as we checked experimentally in preliminary
DDM experiments. We found that a sample made of $20$\% of PBS medium
with bacteria and 80\% of Percoll $\circledR$ was stable upon centrifugation
for 2 hours at 2000 g, which is a time longer than the 30-40 minutes
needed to acquire microscope videos in bright field and dark field.
We note that for large concentrations of bacteria, bacterial aggregation
was found to occur in the presence of Percoll $\circledR$, an effect
presumably attributable to depletion interactions. For this reason
we worked at a low bacterial concentration (about $1\times10^{5}$
bacteria/ml, corresponding to a volume fraction of about $\phi=2\times10^{-7}$),
below the threshold needed to trigger appreciable aggregation during
our experiments. The viscosity of the PBS-Percoll $\circledR$ solution
has been measured by a capillary viscosimeter for different temperatures
in the range $24-25$ \textdegree C. For $T=24$ \textdegree C, the
temperature at which the experiments with bacteria have been performed,
we found $\eta=\left(1.87\pm0.02\right)\,10^{-3}$ $Pa\cdot s$.

\subsubsection*{Polystyrene particles}

We used spherical polystyrene particles (Spherotech (TM) SPHERO (TM)
Biotin Polystyrene Particles), with a certified mean diameter equal
to $0.74$ $\mu m$ (intensity-weighted Nicomp distribution rescaled
to number density). The samples were prepared by gently vortexing
the bottle in order to resuspend the colloidal particles. Serial dilutions
in $15$ $mM$ PBS buffer lead to a final concentration of about $10^{5}$
particles/ml, corresponding to a volume fraction of about $\phi=2.0\times10^{-8}$.
The solution was then sonicated for $10$ min and confined into a
rectangular glass capillary (VITROCOM, internal size: $0.3\times1\times20$
mm) for the microscopy observations. The capillaries were sealed on
both sides with UV glue (UV30-20, Loxeal s.r.l., Cesano Maderno, Italy),
cured for $10$ min under UV light (VL-6.M, Vilbert Lourmat, Marne
la Vall�e, France) and successively loaded on the microscope. The
particles concentration was low enough to ensure that single particles
could be identified and tracked when observed under the microscope,
which allowed a direct-space based characterization of the particles
trajectories, in addition to DDM. The viscosity of the PBS solution
at the temperature $T=24$ \textdegree C at which the experiments
have been performed has been estimated by using literature value for
pure water at the same temperature ($\eta=\left(0.91\pm0.01\right)\,10^{-3}$
$Pa\cdot s$).

\subsection{\label{sec:level2-1-1-2}Differential Dynamic Microscopy}

Microscopy measurements were performed with a Nikon Eclipse Ti-E commercial
microscope equipped with a Hamamatsu Orca Flash 4.0 v2 camera (pixel
size $d_{pix}=6.45$ $\mu$m). Dark-field images are collected with
a $10X$ standard microscopy objective $(\left(NA\right)_{o}=0.15$),
while the sample is illuminated with a condenser stage ($\left(NA\right)_{s}=0.4$)
coupled with a PH3 phase-contrast ring mask. Bright field images are
also collected with the same objective lens and bright field illumination,
by using the same condenser stage with a standard diaphragm. For bacteria
we also used a $40X$ phase-contrast objective ($\left(NA\right)_{o}=0.6$),
the sample being illuminated through the same condenser stage used
in the previous cases, using a proper ring mask in order to achieve
the phase-contrast condition. Each acquisition typically corresponds
to a sequence of $N=50000$ images $I(\mathbf{x},t)$ acquired with
a frame rate $1/\Delta t_{0}$ equal to $100$ fps in the case of
the bacterial suspension and to $20$ fps for the latex particles.
Movies acquired in dark field microscopy exhibit a characteristic
blinking due to the rotation of anisotropic particles, which is not
present in bright field or phase contrast movies (see Supplementary
Movies SM00 and SM01, respectively). To extract quantitative information from these
movies we analyzed them by using the standard DDM algorithm \cite{Croccolo:2006yb,Cerbino:2008if,Giavazzi:2014sj},
which is based on calculating the difference $d(\mathbf{x},t_{0},\Delta t)=I(\mathbf{x},t_{0}+\Delta t)-I(\mathbf{x},t_{0})$
between two images acquired at times $t_{0}$ and $t_{0}+\Delta t$.
Once this quantity is obtained, its spatial Fourier power spectrum
is computed by using a Fast Fourier Transform (FFT) routine and, in
the presence of stationary or quasi-stationary statistical processes,
an average over power spectra with the same $\Delta t$ but different
reference time $t_{0}$ is obtained, which increases the statistical
accuracy of the data. This leads to the so called \textit{image structure
function} (ISF) 
\begin{equation}
d(\mathbf{q},\Delta t)=\left\langle \left|FFT[d(\mathbf{x},t_{0},\Delta t)]\right|^{2}\right\rangle _{t_{0}}\label{eq:struf}
\end{equation}
that captures the dynamics of the sample as a function of the two-dimensional
scattering wave-vector $\mathbf{q}$ and of the delay time $\Delta t$.
The ISF is connected to the (normalized) intermediate scattering function
$f(\mathbf{q},\Delta t)$ \cite{Berne:2000ye} by the relation
\begin{equation}
d(\mathbf{q},\Delta t)=2A(\mathbf{q})\left[1-f(\mathbf{q},\Delta t)\right]+2B(\mathbf{q})\label{eq:strufu}
\end{equation}
where $B(\mathbf{q})$ is a term that accounts for the camera noise,
$A(\mathbf{q})$ is an amplitude term that contains information about
the static scattering from the sample and details about the imaging
system \cite{Giavazzi:2014sj}. The form of the intermediate scattering
function for a sample observed under dark-field imaging, undergoing
both translational and rotational Brownian motion will be discussed
below in Section \ref{sec:level2-1-1-1-1}.

In general, the two-dimensional nature of the ISF provides a powerful
means to probe the sample dynamics along different directions in the
$\mathbf{q}$ plane that may be of particular interest for the problem
under study \cite{Giavazzi:2014fi}. However, whenever the ISF bears
a circular symmetry, as for all the experiment described here, azimuthal
averaging of $d(\mathbf{q},\Delta t)$ is often used to obtain the
one-dimensional function $d(q,\Delta t)$, of the radial wave-vector
$q=\sqrt{q_{x}^{2}+q_{y}^{2}}$ \cite{Cerbino:2008if,Giavazzi:2009xd,Wilson:2011wa,Ferri:2011sh,He:2012bx,Germain:2015gf,Safari:2015lq}.

In order to obtain a reliable determination of the sample dynamics,
which is encoded in the intermediate scattering function, a robust
procedure for estimating the amplitude $A(\mathbf{q})$ and the noise
contribution $B(\mathbf{q})$ in Eq. \ref{eq:strufu} is required.
Dark-field microscopy experiments are very sensitive to non-idealities
such as like dust particles and scratches on the optical elements
and/or on the sample cell, which can give a significant background
signal on top of which the signal from the particle is superimposed.
This background signal appears as an additive, positive term to the
image intensity distribution. The intensity distribution $I(\mathbf{x},t)$
associated with each image can be thus written as the sum of three
independent terms:

\begin{equation}
I(\mathbf{x},t)=I_{0}(\mathbf{x})+I_{s}(\mathbf{x},t)+I_{N}(\mathbf{x},t)\label{eq:ww}
\end{equation}

where $I_{0}(\mathbf{x})$ is a background image (\emph{i.e.} the
intensity distribution that would be observed in absence of the sample),
$I_{s}(\mathbf{x},t)\geq0$ is the contribution to the image from
the particles and $I_{N}(\mathbf{x},t)$ is the camera noise. In our
case, the main contribution to the camera noise is from the shot noise
and we can safely assume that it has zero average: $\left\langle I_{N}\right\rangle =0$
and that it is delta-correlated in both space and time: $\left\langle I_{N}(\mathbf{x+\Delta x},t+\Delta t)I_{N}(\mathbf{x},t)\right\rangle =\left\langle I_{N}^{2}\right\rangle \delta(\Delta x)\delta(\Delta t)$.
If the amplitude of the noise is small compared with the background
intensity $\sqrt{\left\langle I_{N}^{2}\right\rangle }\ll\left\langle I_{0}\right\rangle $
and if the density of the scatterers is low enough to ensure that
only a finite fraction of the image is covered by the particles, an
accurate reconstruction of the background image can be obtained as
$I_{0}(\mathbf{x})=\min_{t}\left\{ I(\mathbf{x},t)\right\} $. In
fact, by picking up for each pixel the lowest intensity value registered
during a suitably large time window - larger than the diffusion time
of a single particle over its image - allows to minimize the additive
contribution from the particles themselves. Once an estimate for $I_{0}(\mathbf{x})$
has been obtained, an estimate of the total amplitude of the fluctuating
parts can be extracted from $C(\mathbf{q})=\left\langle \left|FFT[I(\mathbf{x},t)-I_{0}(x)]\right|^{2}\right\rangle =2A(\mathbf{q})+2B(\mathbf{q})$,
where we have made use of Eq. \ref{eq:ww}. Eq. \ref{eq:strufu} can
be thus rewritten in terms of $C(\mathbf{q})$ as

\begin{equation}
d(\mathbf{q},\Delta t)=2\left[C(\mathbf{q})-B(\mathbf{q})\right]\left[1-f(\mathbf{q},\Delta t)\right]+2B(\mathbf{q}).\label{eq:strufu-1}
\end{equation}

$B(\mathbf{q})$ can be estimated with high accuracy as the intercept
for $\Delta t\rightarrow0$ of $d(\mathbf{q},\Delta t)$. This could
be obtained in practice by fitting $d(\mathbf{q},\Delta t)$ over
a small interval $[0,\,\Delta t_{s}]$ to a polynomial function and
by taking the $0$-th order coefficient. This procedure sets the value
$\left[C(\mathbf{q})-B(\mathbf{q})\right]$ of the amplitude of the
first term, allowing thus a reliable estimate of the relaxation times
in $f(\mathbf{q},\Delta t)$ even if they exceed the width of the
acquisition window. 

\section{Theory\label{sec:Theory}}

In this Section we will first provide a brief summary of the scattering
theory from optically anisotropic particles \cite{Berne:2000ye}.
In the second part we will describe the features of a dark-field imaging
system and the imaging process of anisotropic particles.

\subsection{\label{sec:level2-1-1-1-1}Scattering by anisotropic particles}

\subsubsection*{Small anisotropic particles: Rayleigh scattering}

The description of the scattering of light by a particle much smaller
than the wavelength is usually based on the so-called \textit{Rayleigh
approximation} \cite{Hulst:1957nx}. For simplicity, we will restrict
our discussion to the case of uniaxial particles, whose polarizability
tensor admits a diagonal form with diagonal elements $\alpha_{1}$,
$\alpha_{2}$, $\alpha_{3}$, where $\alpha_{2}=\alpha_{1}$. For
this kind of particles, the anisotropy parameter is defined as $\beta=\alpha_{3}-\alpha_{1}$
and the average (excess) polarizability as $\alpha=\frac{2\alpha_{1}+\alpha_{3}}{3}-V\left(n_{s}^{2}-1\right)$.
Here $V$ is the particle volume and $n_{s}$ is the refractive index
of the dispersion medium. If a plane wave electric
field $\mathbf{A_{0}}(z)=\mathbf{E_{0}}e^{-jkz}$ of wave-number $k$
impinges on such particle, the latter emits a scattered field that,
in addition to the component $E_{VV}$ that is parallel to $\mathbf{E_{0}}$,
also bears a perpendicular component $E_{VH}$. One has \cite{Hulst:1957nx}
\begin{equation}
E_{VV}=S_{VV}\frac{e^{-jkr+jkz}}{jkr}A_{0}(z)\label{eq:vv}
\end{equation}
 and 
\begin{equation}
E_{VH}=S_{VH}\frac{e^{-jkr+jkz}}{jkr}A_{0}(z)\label{eq:vh}
\end{equation}
where $j$ is the imaginary unit, $r$ is the distance from the particle,
and $S_{VV}$ and $S_{VH}$ are dimensionless amplitudes that depend
on the scattering angle $\theta_{s}$ measured with respect to the
direction $z$ of the incident radiation, and on the orientation $(\theta,\phi)$
of the particle. For small scattering angles, the scattering amplitudes
$S_{VV}$ and $S_{VH}$ are given by the following expression \cite{Berne:2000ye}

\begin{equation}
S_{VV}=jk^{3}\alpha+jk^{3}\beta\sqrt{\frac{16\pi}{45}}Y_{2,0}(\theta,\phi),\label{eq:SVV RG}
\end{equation}

\begin{equation}
S_{VH}=jk^{3}\beta\sqrt{\frac{2\pi}{15}}j\left[Y_{2,-1}(\theta,\phi)+Y_{2,1}(\theta,\phi)\right].\label{eq:SVH RG}
\end{equation}

Here $Y_{l,m}(\theta,\phi)$ is the spherical harmonic function of
order $l$, $m$.

Given the shape and the refractive index distribution within a particle,
the calculation of its polarizability tensor is not in general a trivial
task. Even in the Rayleigh regime, closed-form expressions can be
obtained only in presence of particularly simple geometries. In order
to better elucidate how the scattering properties of a particle depend
on its shape and refractive index, it can be worth considering a specific
model system for which a simple analytical description is available,
namely a homogeneous ellipsoid of revolution (spheroid) of semi-axes
$[R,R,\epsilon R]$. In this case, the diagonal elements $\alpha_{j}$
of the (excess) polarizability tensor are given by $\frac{V}{4\pi\alpha_{j}}=L_{j}+\frac{1}{m^{2}-1}$,
where $m=n_{p}/n_{s}$ is the ratio between the refractive index $n_{p}$
of the particle and the refractive index $n_{s}$of the solvent and
$L_{1,2}=L$, $L_{3}=1-2L$, with $L=\epsilon\int_{0}^{\infty}\frac{du}{2\left(u+1\right)^{2}\left(u+\epsilon^{2}\right)^{1/2}}$.
$L$ is a parameter ranging from 0 (flat elliptical disk) to $\frac{1}{2}$
(infinitely long ellipsoid). The value $L=\frac{1}{3}$ corresponds
to the isotropic case. We can also define an anisotropy factor $a=2(3L-1)$
in such a way that $a=0$ corresponds to a sphere, while one has $a=1$
for an infinitely long ellipsoid. If we also define $\delta=m^{2}-1$,
we obtain the following simple expression for the amplitude depolarization
ratio {\cite{Berne:2000ye}

\begin{equation}
\frac{\beta}{\alpha}=\frac{3\delta a}{6+\delta\left(2-a\right)}\simeq\frac{1}{2}\delta a\label{eq:deporatio}
\end{equation}

which in fact is the ratio between the amplitudes of orientation-dependent
component and the orientation-independent component of the scattered
field in Eqs. \ref{eq:vh}, \ref{eq:SVH RG}. It is clear from Eq.
\ref{eq:deporatio} that the contrast in the fluctuation of the depolarized
scattering component depends both on the anisotropy of the particle
and on the refractive index mismatch with respect to the solvent.
Even a strongly anisotropic particle ($a\simeq1$) cannot produce
a significant depolarized signal if it is quasi index-matched with
the solvent ($\delta\simeq0$). We note that, as far as the Rayleigh
approximation is satisfied, the result expressed by Eq. \ref{eq:deporatio}
holds independently of the particle size.

\subsubsection*{Weakly scattering anisotropic particles of larger size: Rayleigh-Gans-Debye
description}

The case of a homogeneous, quasi-index-matched particle can be adequately
described within the \textit{Rayleigh-Gans-Debye (RGD) approximation}
if the overall phase delay associated with the the particle is small:
$|\delta|kd\ll1$ \cite{Hulst:1957nx}. In this approximation, the
amplitude of the wave scattered by the particle can be calculated
as the sum of independent contributions from each portion of particle.
In the RGD approximation the amplitude of the depolarized scattering
is negligible and the scattering amplitude entails only a polarized
component

\begin{equation}
S_{VV}(\mathbf{q})=\frac{j\delta k^{3}}{4\pi}V\cdot F(\mathbf{q})\label{eq:s(q)RGD}
\end{equation}

where

\begin{equation}
F(\mathbf{q})=\frac{1}{V}\int_{V}d^{3}xe^{-j\mathbf{q}\cdot\mathbf{x}}\label{eq:form}
\end{equation}

is the so-called form factor amplitude and $\mathbf{q}=(\mathbf{k_{s}-k_{i}})$
is the transferred momentum, \emph{i.e. }the difference between the
scattering wave-vector $\mathbf{k_{s}}$ and the wave-vector $\mathbf{k_{i}}$
of the incident light. For a particle of a given shape, $F(\mathbf{q})$
has in general an implicit dependence on the particle orientation.
Within the RGD approximation, the scattering in the forward direction
($\mathbf{q}=\mathbf{0}$) does not depend on the details of the particle
shape or orientation, as $F(\mathbf{0})=1$. A non-trivial dependence
of the scattered amplitude on the particle orientation can be observed
only for $q>0$ and if the size of the particle is not too small compared
with the wavelength of light \cite{Berne:2000ye}. 

For a spheroidal particle, such as the one considered in the previous
paragraph, the form factor amplitude can be calculated explicitly:

\begin{equation}
F_{ell}(\mathbf{q})=3\left[\frac{\sin(qR_{\gamma})-qR_{\gamma}\cos(qR_{\gamma})}{(qR_{\gamma})^{3}}\right]\label{eq:f_ell}
\end{equation}

where $R_{\gamma}=R\sqrt{\sin^{2}\theta_{0}+\epsilon^{2}\cos^{2}\theta_{0}}=R\sqrt{1+\left(\epsilon^{2}-1\right)|\mathbf{q\cdot n}|^{2}/q^{2}}$.
Here $\mathbf{n}$ is the unit vector oriented along the particle
axis and $\theta_{0}$ the angle between $\mathbf{n}$ and the transferred
momentum $\mathbf{q}$. In contrast with the forward depolarized scattering
in the Rayleigh regime, here the relative amplitude of the fluctuation
in the scattered intensity due to a rotation of the particle does
not depend at all on the optical contrast, while is strongly influenced
by the overall particle size and by the collection angle, which in
turn determines the transferred momentum $q$.

This property, namely that fact that fluctuation in the intensity
of the light scattered at a large angle by an ensemble of anisotropic
colloidal particles reflects also its rotational dynamics, has been
exploited in DLS for accessing its roto-translational dynamics, as
a complementary approach with respect to low angle depolarized scattering
measurements \cite{Pecora:1968bh,King1973,Schillen1994}.
Eq. \ref{eq:f_ell} takes a particularly simple form if the particle
is not too large compared to the inverse transferred momentum. In
fact, for a relatively small uniaxial anisotropic particle (for which
$qL\lesssim1$, where $L$ is its longer dimension), the expression
in Eq. ,\ref{eq:form}, can be expanded in $(qL)$, leading to:

\begin{equation}
F\simeq1-\frac{q^{2}}{V}\left[I_{zz}+2\left(I_{xx}-I_{zz}\right)\cos^{2}\theta_{0}\right]\label{eq:F_lin}
\end{equation}

or, to the same order in $(qL)$, to

\begin{equation}
P=|F|^{2}\simeq1-\frac{2q^{2}}{V}\left[I_{zz}+2\left(I_{xx}-I_{zz}\right)\cos^{2}\theta_{0}\right],\label{eq:form factor}
\end{equation}
 where $I_{xx}=\int_{V}d^{3}\mathbf{x}\left(x^{2}+z^{2}\right)$ and
$I_{zz}=2\int_{V}d^{3}\mathbf{x}x^{2}$ are the diagonal elements
of the particle tensor of inertia \cite{Berne:2000ye}, and $\theta_{0}$
is the angle between the axis of the particle and the transferred
momentum $\boldsymbol{q}$. The simple harmonic dependence of the
scattered intensity on $\theta_{0}$ described by Eq. \ref{eq:form factor}
will be used in the following paragraphs to link the correlations
of the scattered intensity to the statistical properties of the rotational
motion of the particle.

\subsection{\label{sec:level2-1-1-1}Dark-field microscopy }

\subsubsection*{Dark field imaging}

In the context of optical microscopy, the term dark field (DF) indicates
a family of microscopy configurations characterized by the fact that
the transmitted illumination beam is not collected by the imaging
optics and thus only the light scattered from the sample contributes
to the image. In practice, this can be obtained in a number of different
ways, for example by using dedicated illumination stages. 
\begin{figure*}
\includegraphics[scale=0.4]{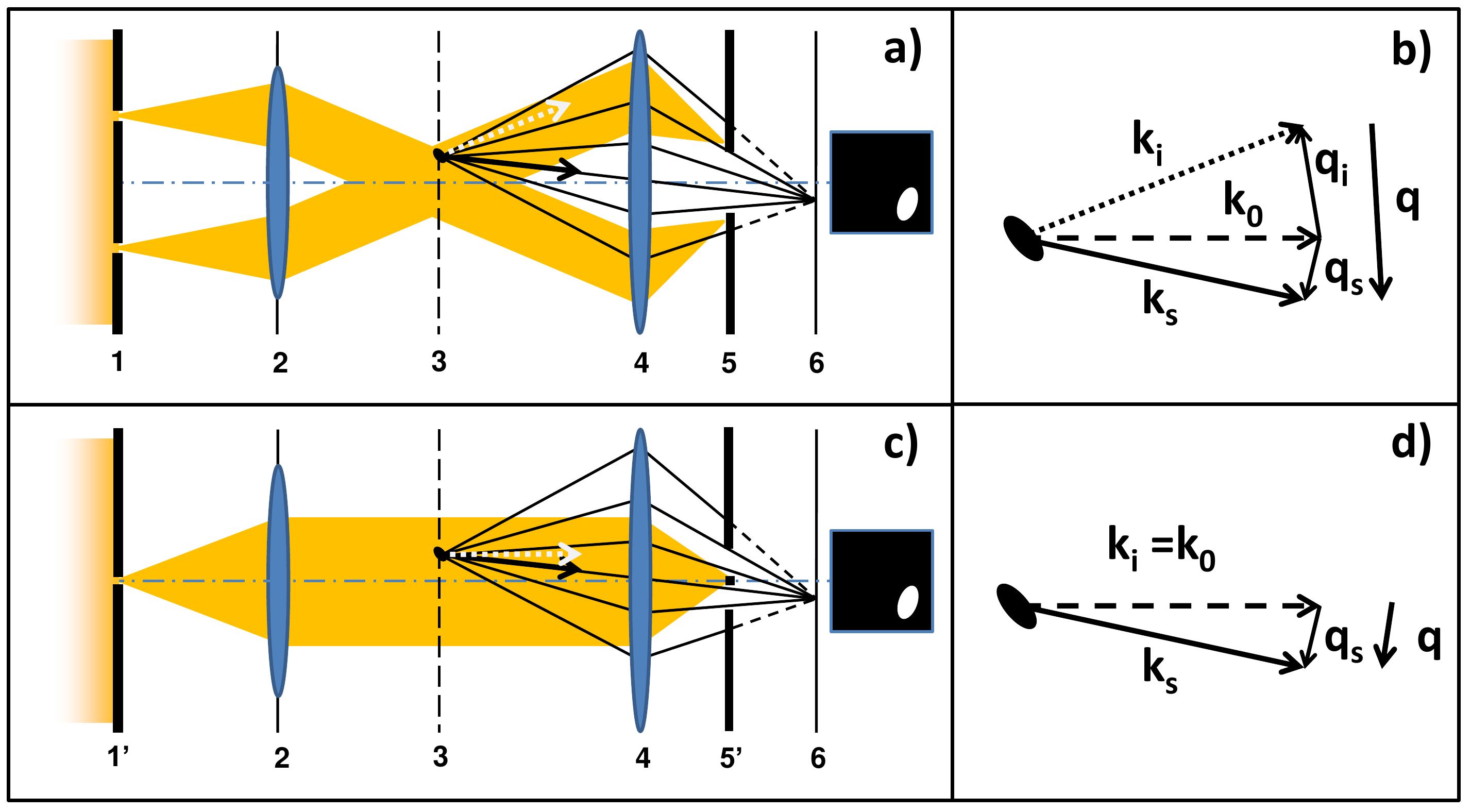}\caption{\label{fig:DF}a) Schematic representation of an incoherent dark field
microscope (IDFM) with Koehler illumination. b) Elementary scattering
process contributing to the image formation in an IDFM. \textbf{$\mathbf{k}_{i}$}
denotes the incident wave-vector, \textbf{$\mathbf{k}_{s}$} is the
scattering wave-vector, $\mathbf{q}=\boldsymbol{k}_{s}-\boldsymbol{k}_{i}$
is the transferred momentum, and \textbf{$\mathbf{k}_{0}$} the wave-vector
oriented along the optical axis. The transferred momentum can be also
written as $\boldsymbol{q}=\boldsymbol{q}_{s}-\boldsymbol{q}_{i}$,
where $\boldsymbol{q}_{s}=\boldsymbol{k}_{s}-\boldsymbol{k}_{0}$
and $\boldsymbol{q}_{i}=\boldsymbol{k}_{i}-\boldsymbol{k}_{0}$. c)
Schematic representation of a coherent dark field microscope (CDFM)
in a custom bench-top homodyne near-field scattering set-up (see text
for additional details). d) Scattering process in CDFM. The incident
wave-vector \textbf{$\mathbf{k}_{i}$} is parallel to optical axis.
The transferred momentum $\boldsymbol{q}$ is given by $\boldsymbol{k}_{s}-\boldsymbol{k}_{i}$,
where $\boldsymbol{k}_{s}$ is the scattering wave vector. }
\end{figure*}
A common implementation, which is compatible with most commercial
inverted microscopes is the one reported in Fig. \ref{fig:DF} a),
where a schematic representation of a microscope with Koehler illumination
is shown. In this configuration, a circular aperture or radius $R_{ring}$
carved into an opaque mask is placed in the back focal plane of the
condenser lens, of focal length $\left(FL\right)_{cond}$. In this
condition the object plane is illuminated only by rays forming an
angle $\theta\simeq\arcsin(NA)_{c}$, where $\left(NA\right)_{c}=\arctan\left[R_{ring}/\left(FL\right)_{cond}\right]$.
The dark field condition is achieved by using an objective lens with
numerical aperture $\left(NA\right)_{o}<\left(NA\right)_{c}$, so
that the transmitted beam is not collected by the objective. In practice,
this is commonly achieved by coupling a high numerical aperture phase
contrast ring with a low power objective. In this configuration, the
illumination beam can be thought of as the incoherent superposition
of many plane waves propagating at angles $\theta\simeq\arcsin\left(NA\right)_{c}$
with respect to optical axis \emph{i.e.} the sample is illuminated
by a collection of uncorrelated coherent patches of transverse size
$\Lambda_{s}\simeq\lambda/(NA)_{c}$. Since $\left(NA\right)_{o}<\left(NA\right)_{c}$,
the size of these coherent patches is smaller than the transverse
size $\Lambda_{o}\simeq\lambda/\left(NA\right)_{o}$ of the objective
PSF. For this reason, the imaging process is to all effects incoherent,
as interference effects between different scattering centers are negligible.
For this reason, we refer to this configuration as Incoherent Dark-Field
Microscopy (IDFM).

A different implementation of a dark field imaging system is the one
shown in \ref{fig:DF} c), which is typically adopted in homodyne
imaging and/or near-field scattering set-ups \cite{Giglio:2000_NFS,Cerbino:2009ta}.
In this case the sample is illuminated by a highly coherent beam and
the transmitted light is blocked by a small opaque patch placed in
the back focal plane of the objective lens. We call this configuration
Coherent Dark Field Microscopy (CDFM), since the sample is ideally
illuminated by a single plane wave propagating along the optical axis
and the light scattered by different points of the sample always bears
a well defined relative phase. The superposition of the scattering
patterns from different points of the sample takes thus place on a
coherent basis, since interference effects cannot be neglected and
the imaging process is linear in the complex amplitude rather than
in the intensity, as it is for IDFM \cite{ob:goodmanfourier}.

According to the nomenclature introduced in \cite{Giavazzi:2009xd},
IDFM is thus a \emph{linear} imaging system, whereas CDFM is not,
in that the image intensity recorded in presence of two particles
is not given by just the sum of the intensities associated with the
two particles imaged separately. For this reason, IDFM is to be preferred
for Digital Fourier Microscopy experiments, where a \emph{linear}
space-invariant imaging system turns out to beneficial \cite{Giavazzi:2014sj}.
We note that, as pointed out in Ref. \cite{Bayles:2016_darkDDM},
dark field imaging can in some cases lead to a non-homogeneous illumination
pattern, which violates the space-invariant assumption. In our optical
set-up this effect was negligible, in that the illumination was found
to be uniform across the entire field of view. 

\subsubsection*{The incoherent dark field microscope as a fixed, non-zero angle scattering
set-up}

In a dark field microscope the main contribution to the particle image
intensity comes from light scattered at a non-zero angle, roughly
corresponding to the illumination numerical aperture. As a consequence,
if the particle is anisotropic, its reorientation produces a modulation
in the image intensity. This property opens to the possibility of
exploiting IDFM to study the rotational dynamics of anisotropic particles.
The main elements of typical dark field microscopy setup are shown
in the simplified scheme in Fig. \ref{fig:DF}. The elementary scattering
process contributing to the image formation is shown in Fig. \ref{fig:DF}
b), where \textbf{$\mathbf{k}_{i}$} denotes the incident wave-vector,
\textbf{$\mathbf{k}_{s}$} is the scattering wave-vector and $\mathbf{q}=\mathbf{k}_{s}-\mathbf{k}_{i}$
is the transferred momentum. By introducing $\mathbf{k}_{0}$ as the
wave-vector oriented along the optical axis, having the same amplitude
of \textbf{$\mathbf{k}_{i}$} and \textbf{$\mathbf{k}_{s}$} we can
express the transferred wave-vector as $\mathbf{q}=\mathbf{q}_{s}-\mathbf{q}_{i}$,
where $\mathbf{q}_{s}=\mathbf{k}_{s}-\mathbf{k}_{0}$ and $\mathbf{q}_{i}=\mathbf{k}_{i}-\mathbf{k}_{0}$.
In the image plane, the total intensity $I$ associated with the particle's
image can be calculated as the (incoherent) sum of all such processes

\begin{equation}
I_{p}=V\int d\mathbf{q}_{i}\int d\mathbf{q}_{s}I_{0}(\mathbf{q}_{0})P(\mathbf{q}_{s}-\mathbf{q}_{i})T(\mathbf{q}_{s})\label{eq:Intensity}
\end{equation}

where the function $I_{0}(\mathbf{q}_{0})$, representative of the
angular distribution of the illumination beam, weights the contribution
of each incoming plane wave, the particle form factor $P(\mathbf{q})$
accounts for the scattering properties of the sample, and the incoherent
transfer function $T(\mathbf{q}_{s})$ quantifies the collection efficiency
of the objective lens. 

Our choice of a thin phase contrast ring of numerical aperture $(NA)_{c}$
makes the evaluation of the integral on the right hand side of Eq.
\ref{eq:Intensity} particularly simple, as $I_{0}(\mathbf{q}_{0})\simeq I_{0}\delta(|\mathbf{q}_{0}|-q^{*})$,
where $q^{*}=2k_{0}\sin(\frac{\theta_{c}}{2})\simeq k_{0}\sin(\theta_{c})=k_{0}\left(NA\right)_{c}$.
In the general, $T(\mathbf{q})$ is an azimuthally symmetric function
of width $\sim k_{0}\left(NA\right)_{o}<k_{0}\left(NA\right)_{c}$
centered around $\mathbf{q}=0$. If we take the (rather crude) approximation:
$T(\mathbf{q}_{s})\simeq\delta(\mathbf{q}_{s})$, corresponding to
considering only the light scattered along the optical axis, we obtain
the following simple result, where the total intensity of the particle's
image is written as an azimuthal average of the particle form factor
performed for $|\mathbf{q}|=q^{*}$: $I_{p}\propto\int d\mathbf{q}\delta(|\mathbf{q}|-q^{*})P(\mathbf{q})=\int_{0}^{2\pi}d\alpha P\left(\left[\begin{array}{ccc}
q_{\perp}^{*}\cos\alpha, & q_{\perp}^{*},\sin\alpha, & q_{\parallel}^{*}\end{array}\right]\right)$, where $q_{\perp}^{*}\simeq q^{*}$ and $q_{\parallel}^{*}\simeq0$
are the projections of the transferred momentum in the direction perpendicular
and parallel to the optical axis, respectively . In IDFM, the main
contribution to the particle image intensity comes from light scattered
at a non-zero angle, roughly corresponding to the illumination numerical
aperture. If we assume for the form factor $P$ the expression given
in Eq. \ref{eq:form factor}, $I_{p}$ can be easily integrated leading
to:

\begin{equation}
I_{p}\propto1-\frac{2q^{*2}}{V}\left[I_{xx}+\left(I_{zz}-I_{xx}\right)\sin^{2}(\theta)\right]\label{eq:form factor-1}
\end{equation}

or, equivalently

\begin{equation}
I_{p}=I_{p,0}\left[1+cY_{20}(\theta)\right]\label{eq:IP}
\end{equation}

where $Y_{20}(\theta)=\sqrt{\frac{5}{16\pi}}\left(3\cos^{2}\theta-1\right)$
is the spherical harmonic of order 2,0, $\theta$ is the angle between
the axis of the particle and the optical axis, $I_{p,0}$ is a constant
amplitude and $c$ is a constant. Eq. \ref{eq:IP} shows that the
intensity signal due to an anisotropic particle depends on the particle
orientation with respect to the optical axis, which explains why a
particle that undergoes a rotational Brownian motion appears as a
randomly blinking object. We note that considering a finite value
for the objective numerical aperture does not affect this result.
In fact, in that case $I_{p}$ is still given by the integral of $P(q)$
over an annular region of radius $q^{*}$ but with a finite width
$\sim k_{0}\left(NA\right)_{o}$, which does not affect the angular
dependence of $I_{p}$ in Eq. \ref{eq:IP} but only the prefactor
$c$. We note that Eq. \ref{eq:IP} holds under the same hypotheses
under which Eq. \ref{eq:form factor} is valid, namely that the $Lq^{*}\lesssim1$,
where $L$ is longer dimension of the particle. For very large particles
a more complex expression is expected, involving higher order spherical
harmonics.

\subsubsection*{Dark field DDM probes the roto-translational dynamics of anisotropic
particles}

The fact that the image intensity corresponding to one particle depends
on the particle orientation can be in principle used to assess the
rotational dynamics by studying the characteristic time of blinking
in movies obtained by IDFM (see Supplementary Movie SM01 and SM02).
For the case of interest in this work,\emph{ i.e. }when the particle
undergoes a rotational Brownian motion, its rotational diffusion coefficient
$D_{R}$ can be obtained by calculating the intensity temporal auto-correlation
function

\begin{equation}
C_{p}(\Delta t)=\left\langle I_{p}(t+\Delta t)I_{p}(t)\right\rangle =I_{p,0}^{2}\left(1+c^{2}e^{-6D_{R}t}\right),\label{eq:rotainte}
\end{equation}

whose characteristic time $(6D_{R})^{-1}$ would give immediate access
to the rotational diffusion coefficient $D_{R}$ \cite{Berne:2000ye}.
However, measuring $C_{p}(\Delta t)$ is usually quite difficult:
for instance, particles can disappear from the image when they exit
the focal region or two particles might superimpose along the optical
axis and give rise to spurious effects. These difficulties can be
bypassed by working in the wave-vector space, as done in Dark Field
Differential Dynamic Microscopy (d-DDM)\cite{Bayles:2016_darkDDM}.
The image intensity distribution of a collection of identical particles
observed in IDFM is given by

\begin{equation}
I_{s}(\mathbf{x},t)=\sum_{n}I_{p}^{(n)}(t)\psi(\mathbf{x}-\mathbf{x}_{n})\label{eq:collint}
\end{equation}

where $\psi(\mathbf{x})$ is a function that describes the distribution
of the individual scatterers in real space. If $\hat{I_{s}}(\mathbf{q},t)$
is the spatial Fourier transform of $I_{s}(\mathbf{x},t)$, an alternative
quantification of the particle dynamics is provided by the \textsl{intermediate
scattering function}
\begin{equation}
\left\langle \hat{I_{s}}(q,t+\Delta t)\hat{I_{s}}^{*}(q,t)\right\rangle =N|\psi(q)|^{2}C_{p}(\Delta t)e^{-D_{T}q^{2}\Delta t}\label{eq:unno}
\end{equation}

where $D_{T}$ is the translational diffusion coefficient of the particles.
Eq. \ref{eq:unno} can be rewritten by introducing the \textsl{normalized
intermediate scattering function \cite{Berne:2000ye}}

\begin{equation}
f(\mathbf{q},\Delta t)=\frac{\left\langle \hat{I_{s}}(q,t+\Delta t)\hat{I_{s}}^{*}(q,t)\right\rangle }{\left\langle |I(q,t)|^{2}\right\rangle }=\alpha e^{-\Gamma_{1}(q)\Delta t}+(1-\alpha)e^{-\Gamma_{2}(q)\Delta t},\label{eq:normalisf}
\end{equation}

where the contribution of the two relaxation modes with decay rates
$\Gamma_{1}=6D_{R}+D_{T}q^{2}$ and $\Gamma_{2}=D_{T}q^{2}$ can be
appreciated. This form for the intermediate scattering function is
pretty common, as is found also in D-DLS and p-DDM experiments \cite{Piazza:1990fr,Giavazzi:2016_jpcm}.

\section{Results and discussion}

In this Section, we demonstrate d-DDM as an effective powerful tool
for the simultaneous determination of the translational and rotational
dynamics by presenting results obtained with two representative samples:
a suspension of quasi-index matched anisotropic colloidal particles
(non-motile rod-shaped bacteria) and a suspension of quasi-monodisperse
spherical colloids both in a non-aggregated state and forming small
clusters.

\subsection{\label{sec:level2-1-4-2}Bacteria}

A first set of measurements was performed on a suspension of density-matched
non-motile coliform bacteria, prepared as described in the Material
and Methods Section. Movies of the sample are acquired with dark field,
bright field and phase contrast microscopy. A representative dark field
image of the sample is shown in Fig. \ref{fig:isf-1} (see also Supplementary
Movie SM01): the bacterial particles appear as bright spots diffusing
on a dark background and whose intensity fluctuates with a characteristic
time of about $1$ $s$. We will show that these two processes, namely,
the concentration fluctuation due to the translational motion of the
particles and the intensity fluctuations caused by the rotational
dynamics of single particles, are well captured by the d-DDM analysis. 

\begin{figure}
\includegraphics[width=0.8\columnwidth]{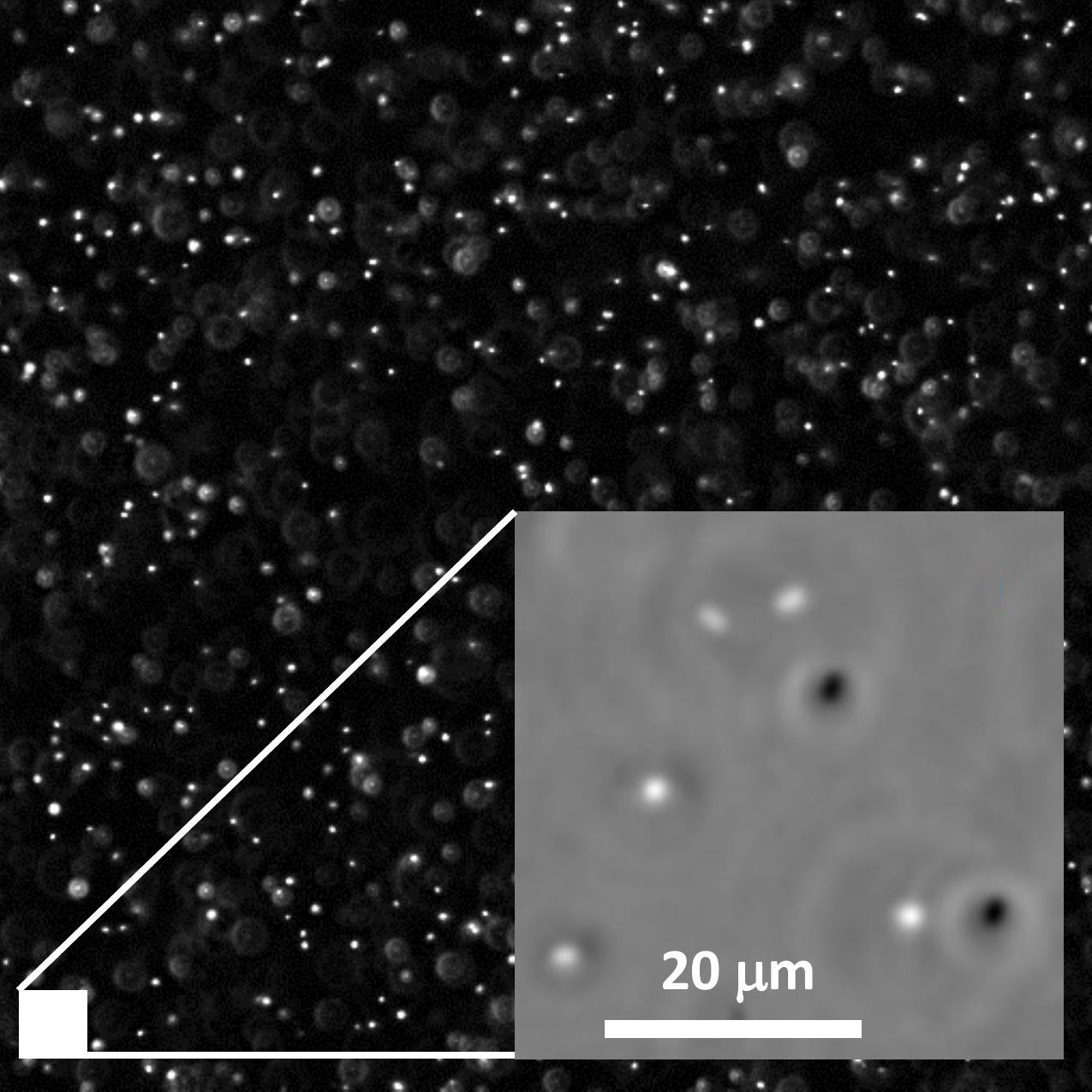}\caption{Main panel: representative dark field image of the bacterial suspension,
where the optical background has been subtracted as described in the
Material and Methods Section. Inset: representative phase-contrast
image of the same sample. \label{fig:isf-1}}
\end{figure}

In Fig. \ref{fig:isf} we show for some representative values of the
wave-vector $q$, logarithmically spaced in the in range $[0.075,\,2.3]$
$\mu m^{-1}$, the intermediate scattering functions obtained with
d-DDM analysis. As expected from Eq. \ref{eq:normalisf}, the observed
relaxation exhibits two distinct decays. This is particularly evident
in the low-$q$ regime, where the time-scale separation between the
two decays is more pronounced (see Fig. \ref{fig:isf} b). The intermediate
scattering functions are well described by a sum of two simple-exponential
decays (continuous lines in Fig. \ref{fig:isf}). Fitting the obtained
curves to Eq. \ref{eq:normalisf} provides thus an estimate for the
two $q$-dependent relaxation rates $\Gamma_{1}(q)$ and $\Gamma_{2}(q)$.
The so-obtained $\Gamma_{1}$ and $\Gamma_{2}$ are shown in Fig.
\ref{fig:rates} together with the best fitting curves: $\Gamma_{1}(q)=D_{T}q^{2}+6D_{R}$,
and $\Gamma_{2}(q)=D_{T}q^{2}$. From this last fit we obtain an estimate
for the translational and the rotational diffusion coefficient of
the bacteria: $D_{T}=\left(1.57\pm0.02\right)\cdot10^{-1}$ $\mu m^{2}s^{-1}$
and $D_{R}=\left(1.5\pm0.1\right)\cdot10^{-1}$ $s^{-1}$.

\begin{figure*}
\includegraphics[width=0.8\paperwidth]{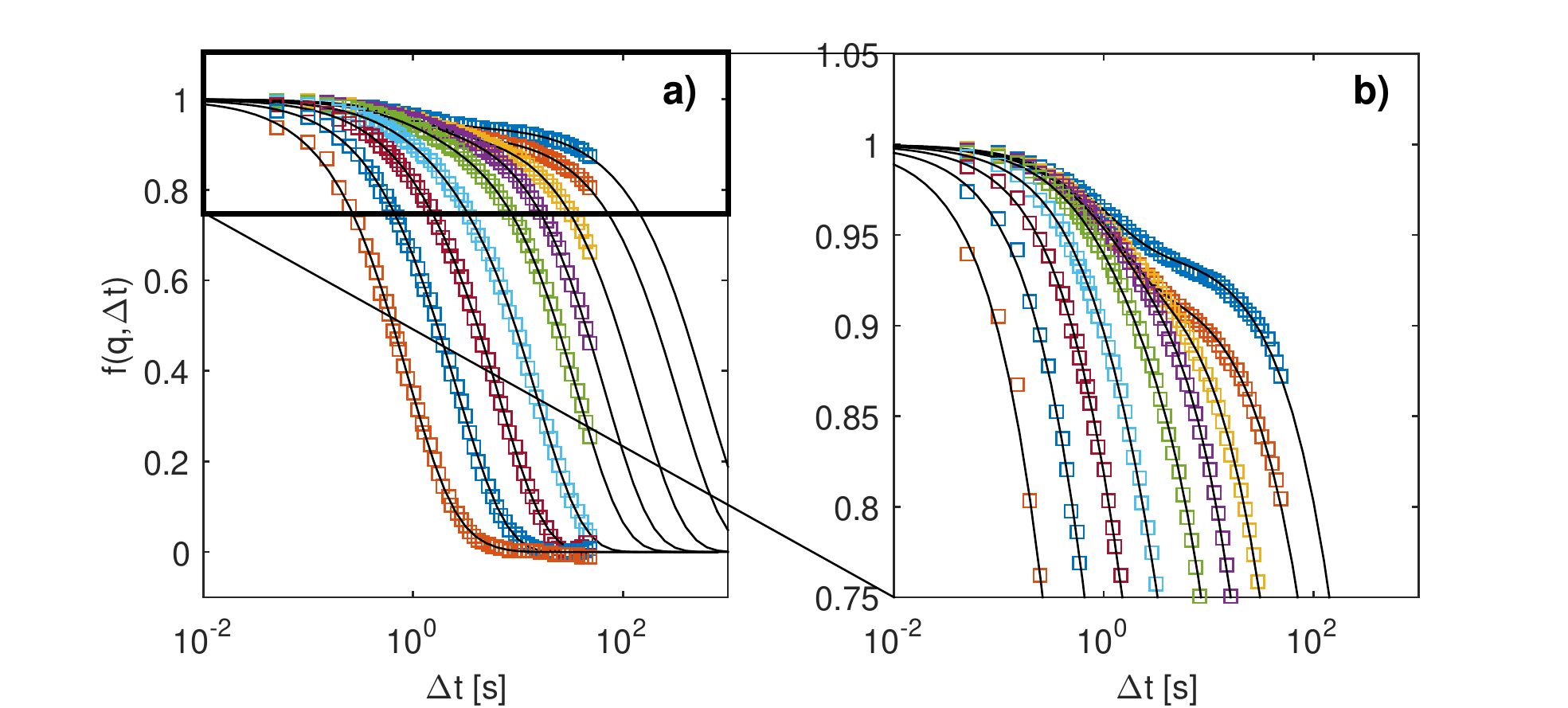}\caption{Symbols: intermediate scattering function $f(q,\Delta t)$ obtained
for different wave vectors $q=0.075$, $0.11$,$0.19$, $0.30$, $0.44$,
$0.65$, $0.98$, $1.5$, $2.3$ $\mu m^{-1}$ in d-DDM experiments.
Continuous lines: best fitting curves with the model function given
in Eq. \ref{eq:normalisf}. \label{fig:isf}}
\end{figure*}
The simultaneous measurement of both $D_{T}$ and $D_{R}$ allows
estimating the size of the bacterial particles. We consider two simple
models, for which analytical expressions for the diffusion coefficients
are available, namely a spheroid of semiaxis $\bar{R}\epsilon^{-1/3}\left(1,1,\epsilon\right)$
\cite{KOENIG} and a cylinder of length $L$ and radius $r$ \cite{TIRADO1984,PhysRevE.50.1232_lowen}.
Explicit expression for $D_{R}$ and $D_{T}$ in the two cases are
reported in Appendix A and can be inverted numerically to determine
the best estimates for the geometrical parameters in the two cases:
$\bar{R}=0.70\pm0.02$ $\mu m$, $\epsilon=2.3\pm0.3$ for the spheroid
and $L=1.6\pm0.2$ $\mu m$, $r=0.6\pm0.2$ $\mu m$ for the cylinder.
We note that both models provide meaningful results, that are in good
relative agreement and are fully compatible with both literature values
\cite{Grossman1982,Nelson2000} and direct, high magnification microscopy
observations (see Fig. \ref{fig:isf-1}).

As a consistency check, we also analyzed bright field and phase-contrast
movies of the same sample. In both cases, the obtained intermediate
scattering functions show a single decay that is very well fitted
by a simple exponential function (data not shown). The corresponding
$q$-dependent relaxation rates are reported in Fig. \ref{fig:rates}
and exhibit a very clean $q^{2}$ scaling and the estimated translational
diffusion coefficient $D_{T}=\left(1.58\pm0.01\right)$ $\mu m^{2}/s$
is in excellent agreement with the results of d-DD. Of note, the wave-vector
range over which a reliable estimate of the dynamics can be obtained
is not the same for all the imaging conditions. In particular, we
observe that, although dark field and bright field measurements are
performed with the same objective, frame rate and duration of the
acquisition, dark field DDM is much more effective in probing the
dynamics in the the low-$q$ regime. This can explained by inspecting
the static scattering amplitudes $A(q)$ obtained from the DDM analysis
in the two datasets show in the nset of Fig. \ref{fig:rates}, which
outline the effect of the different transfer function of the two methods.
The bright field amplitude shows the characteristic depression at
low $q$, reflecting the fact that bacteria, being quasi-index matched
with the solvent, behave as phase objects \cite{Giavazzi:2009xd}.
On the contrary, the dark field amplitude is a monotonically decreasing
function of $q$. The phase contrast data, being obtained with a different
microscope objective, with larger magnification and numerical aperture
(40 vs 10 and 0.6 vs 0.15), cover a $q$ range shifted by approximately
half a decade to larger wave-vectors.

\begin{figure}
\includegraphics[width=0.8\columnwidth]{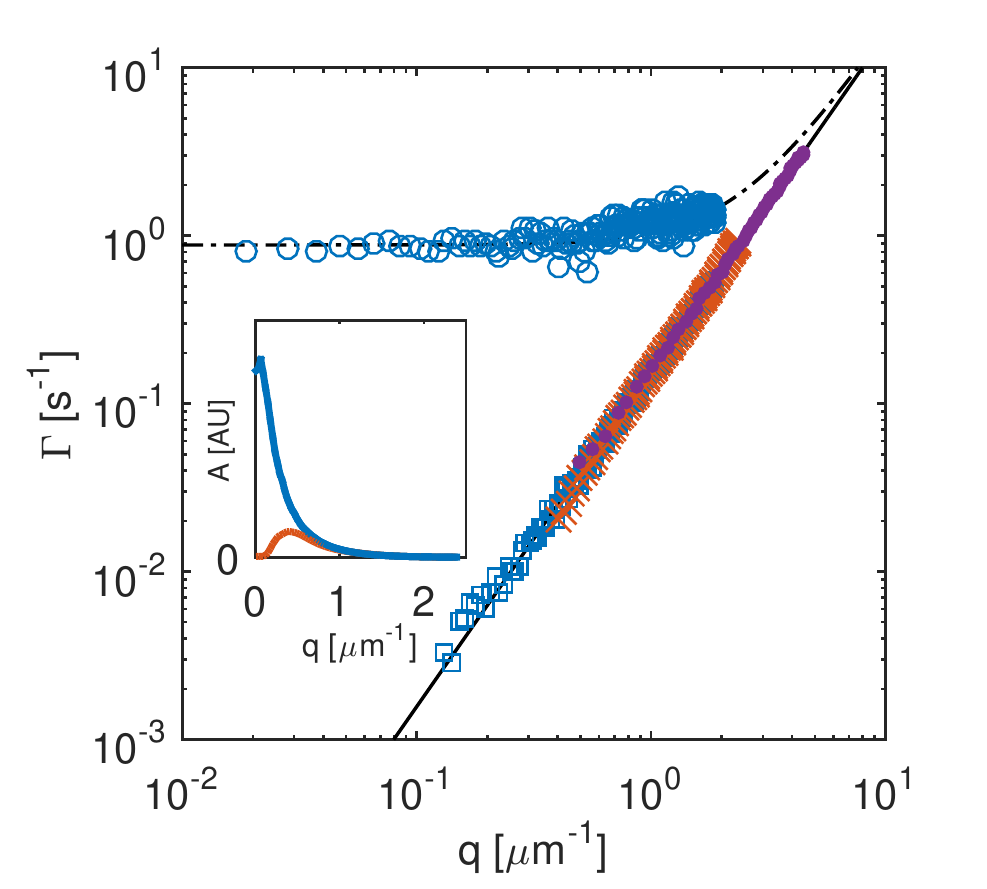}

\caption{Decorrelation rates as a function of the wave-vector $q$ obtained
from the fit of the intermediate scattering functions obtained in
dark field (open blue circles and squares), in bright field (orange
crosses) and in phase-contrast (purple circles) microscopy experiments
on a dilute bacterial dispersion. The dash-dotted and the continuous
lines are the best fits to the curves $\Gamma_{1}(q)=6D_{R}+D_{T}q^{2}$
and $\Gamma_{2}(q)=D_{T}q^{2}$, respectively, which lead to the estimates
$D_{T}=\left(1.57\pm0.02\right)\cdot10^{-1}$ $\mu m^{2}s^{-1}$ and
$D_{R}=\left(1.5\pm0.1\right)\cdot10^{-1}$ $s^{-1}$. Bright field
and phase-contrast data provided the value $D_{T}=\left(1.58\pm0.01\right)$
$\mu m^{2}/s$, obtained by fitting the experimental data to the curve
$\Gamma_{2}(q)=D_{T}q^{2}$ (fitting curve not shown). Inset: static
amplitude $A(q)$ obtained with the same microscope objective from
DDM analysis of bright field (red squares) and dark field (blue circles)
movies of the same bacterial dispersion. While the dark field amplitude
is a decreasing function $q$, the optical transfer function for the
bright field experiments is characterized by a visible depression
at small scattering wave-vectors. \label{fig:rates}}
\end{figure}

\subsection{\label{sec:level2-1-4-2-1}Spherical colloids}

A second set of d-DDM measurements was performed on a suspension of
monodisperse spherical latex particles. A first sample (Sample 1)
was prepared with a 5 minutes sonication stage before the measurement,
as described in the Materials and Methods Section. Surprisingly, we
found that d-DDM analysis showed two distinct decays $\Gamma_{1}(q)$
and $\Gamma_{2}(q)$ for the ISF (\ref{fig:AMPS-1}), as previously
found for the bacteria. This finding contrasted our expectation to
observe a single relaxation mode due to translational diffusion of
the particles. In order to better understand the reason of such unexpected
behavior we thus performed a bright field DDM experiment on the same
sample: as expected for a reasonably monodisperse sample, the experimental
ISFs were well fitted to a single exponential decay and the so-obtained
relaxation rate $\Gamma(q)$ was well fitted to the function $\Gamma(q)=D_{T}q^{2}$
(Fig. \ref{fig:AMPS-1}). The estimate $D_{T}^{BF}=0.475\pm0.05$
\textmu m$^{2}$/s for the translational diffusion coefficient of
the particles is fully compatible with the value found in dark field
experiments, which gave $D_{T}^{DF}=0.49\pm0.01$ \textmu m$^{2}$/s
and $D_{R}=1.3\pm0.05$ $s^{-1}$ for the translational and rotational
diffusion coefficients, respectively. However, none of them was found
to be compatible with the value $R_{H}^{cert}=0.37\pm0.2$ $\mu m$,
certified by the producer for the hydrodynamic radius of the particles.
Both values $R_{H}^{BF}=50.3\pm0.05$ $\mu m$ and $R_{H}^{DF}=49\pm1$
$\mu m$, obtained from $D_{T}^{BF}$ and $D_{T}^{DF}$, respectively,
\emph{via} the Stokes-Einstein relation are about $30\%$ larger than
the nominal value. We interpreted all these results as consequences
of the presence of anisotropic particles in suspension, originated
from aggregation of the spherical particles. Careful inspection of
dark field movies (Supplementary Movie SM02) further supported this
hypothesis in that it pointed to the presence of a small number of
blinking particles. 

To obtain final confirmation, we performed measurements on a second
sample (Sample 2) that was carefully prepared from the same batch
of particles by using a longer sonication stage ($30$ minutes instead
of $5$). With Sample 2, the results of DDM analysis of bright field
and dark field movies (see Supplementary Movie SM03) confirm the absence
of aggregates. In both cases, the ISFs exhibit a \emph{single} exponential
decay, with a relaxation rate displaying a clean quadratic scaling
with $q$, from which we obtain the estimate $D_{T}=0.695\pm0.01$
\textmu m$^{2}$/s for the translational diffusion coefficient (Fig.
\ref{fig:AMPS-1}). The extracted hydrodynamic radius $R_{H}=35\pm1$
$\mu m$ is now fully compatible with the nominal size of the particle,
as certified by the producer. We could therefore safely conclude that
Sample 1 contained aggregated particles and that d-DDM provides a
very powerful means to spot the presence of small aggregates in colloidal
dispersions of spherical particles. In particular, when sizing unknown
samples, it would be recommended to complement bright field DDM experiments
with dark field ones, to check whether a rotational diffusion decay
mode due to aggregates is present or not.

We then turned to assessing how far can d-DDM be brought in obtaining
information on the aggregates. To this aim we evaluated the values
expected for the translational and rotational diffusion coefficients
of small clusters of particles (dimers, trimers, tetramers, etc.){}
\cite{hoffmann2009}. In particular, the value $D^{cert}{}_{T}=0.65$
\textmu m$^{2}$/s, corresponding to the hydrodynamic radius certified
by the producer, gives the following values for small clusters: $D_{T}^{di}=0.48$
\textmu m$^{2}$/s (dimers), $D_{T}^{tri}=0.41$ \textmu m$^{2}$/s
(trimers), and $D_{T}^{tetra}=0.38$ \textmu m$^{2}$/s (tetramers).
We note that these values are very close to each other, which explains
why translational diffusion does not discriminate very effectively
between these contributions. This is confirmed by our d-DDM results,
as we found that the translational diffusion coefficient extracted
with d-DDM exhibited an intermediate value between the expectation
for monomers and for the small clusters. By contrast, the rotational
diffusion coefficient is not sensitive to the presence of monomers.
Consistently, the value obtained with d-DDM ($D_{R}=1.3\pm0.05$ $s^{-1}$)
was found to be within the range of the rotational diffusion coefficient
expected for small clusters:$D_{R}^{di}=1.39$ $s^{-1}$
(dimers), $D_{T}^{tri}=0.87$ $s^{-1}$ (trimers), and $D_{R}^{tetra}=0.67$
$s^{-1}$ (tetramers).

\begin{figure}
\includegraphics[width=0.8\columnwidth]{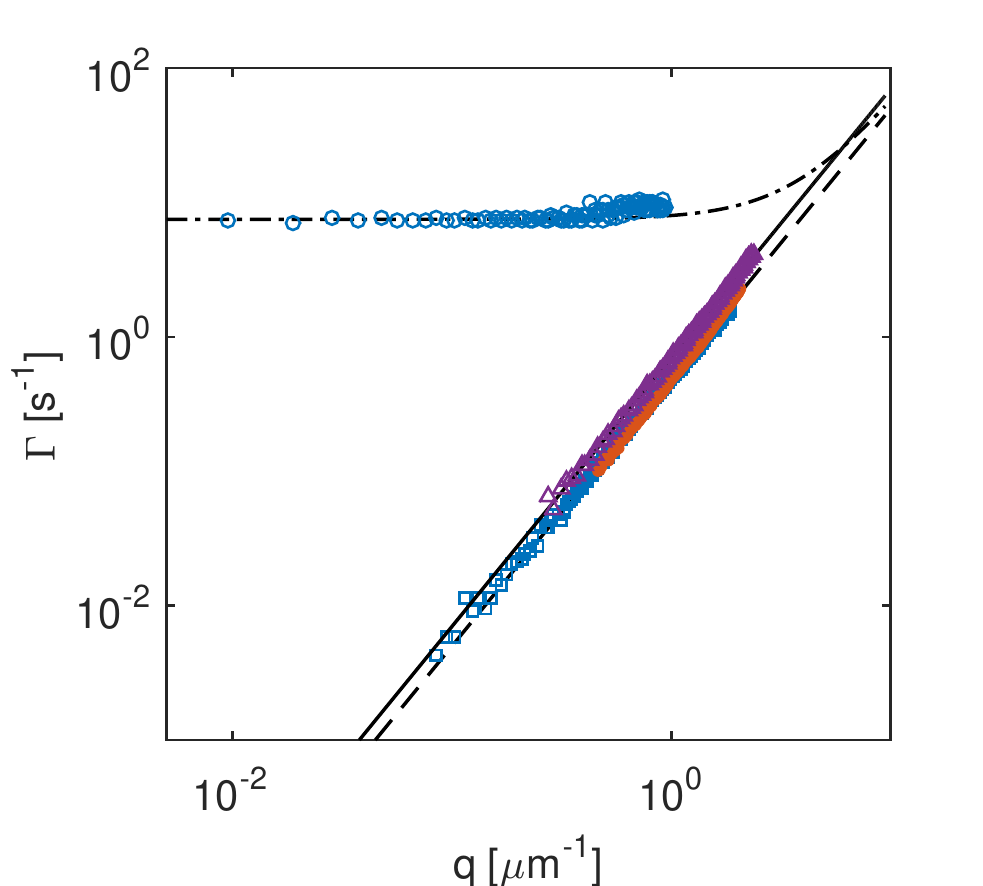}\caption{Decorrelation rates obtained from the fit of the intermediate scattering
functions obtained in dark field (open blue circles and squares) and
in bright field (orange dots) microscopy experiments on a partially
aggregated suspension of spherical colloidal particles (Sample 1).
The dash-dotted and the dashed lines are the best fits to the curves
$\Gamma_{1}(q)=6D_{R}+D_{T}^{BF}q^{2}$ and $\Gamma_{2}(q)=D_{T}^{BF}q^{2}$,
respectively, which lead to the estimates $D_{T}^{DF}=0.49\pm0.02$
$\mu m^{2}s^{-1}$ and $D_{R}=1.3\pm0.05$ $s^{-1}$. Bright field
data provided the value $D_{T}^{BF}=0.475\pm0.05$ $\mu m^{2}s^{-1}$,
obtained by fitting the experimental data to the curve $\Gamma_{2}(q)=D_{T}^{BF}q^{2}$
(fitting curve not shown). Upward purple triangles correspond to the
single decorrelation rate $\Gamma(q)$ measured in a dark field microscopy
experiment on a non-aggregated suspension of the same particles (Sample
2), The continous line is the best fit to the curve $\Gamma(q)=D_{T}q^{2}$,
leading to the estimate $D_{T}=0.695\pm0.01$ $\mu m^{2}s^{-1}$.
\label{fig:AMPS-1}}
\end{figure}

In order to check the consistency of this picture, we performed a
detailed Video Particle Tracking (VPT) analysis on the same movies,
which is very delicate and time consuming but provides a more detailed
information about the sub-populations of which the sample is composed.
Particles trajectories are obtained with the Particle Tracker Plugin,
included in the Mosaic Suite for ImageJ. With a custom software written
in Matlab, we extracted from each trajectory the particles mean squared
displacement and, by fitting the resulting curve as a function of
the time delay we obtain and estimate for the translational diffusion
coefficient $D_{T}$ of each single particle. By rejecting trajectories
shorter than $200$ time steps, we obtained $D_{T}$ for about $1000$
particles for each sample. In Fig. \ref{fig:AMPS-1-1}, we report
the histograms representing the distributions of the values of $D_{T}$
obtained for each of the two samples. For Sample 2, the distribution
of $D_{T}$ is fairly symmetric and it is well described by a Gaussian
function with mean value $\bar{D}_{T}=0.69$ $\mu m^{2}/s$ and standard
deviation $\sigma_{D_{T}}=0.1$ $\mu m^{2}/s$. This result is compatible
with a moderately dispersed distribution peaked around a value that
is in excellent agreement with estimated $D_{T}=0.695\pm0.01$ \textmu m$^{2}$/s
obtained for the translational diffusion coefficient from the d-DDM
analysis. For Sample 1, the same SPT-based analysis provides a completely
different result. The distribution of $D_{T}$ is broader and, beside
a peak centered about a value compatible with $\bar{D}_{T}$ , a secondary
peak for $D_{T}\simeq0.4$ $\mu m^{2}/s$ is also clearly visible.
As shown in Fig. \ref{fig:AMPS-1-1}, the range covered by this secondary
peak is compatible with the translational diffusion coefficients expected
for small clusters of monomers, assuming $\bar{D}_{T}$ as the diffusion
coefficient of a monomer. These results strongly corroborate our interpretation
and confirm that d-DDM can be used to measure with high sensitivity
the translational dynamics of spherical colloids and to spot aggregation
of spherical particles in a simple and effective way. 

\begin{figure}
\includegraphics[width=0.8\columnwidth]{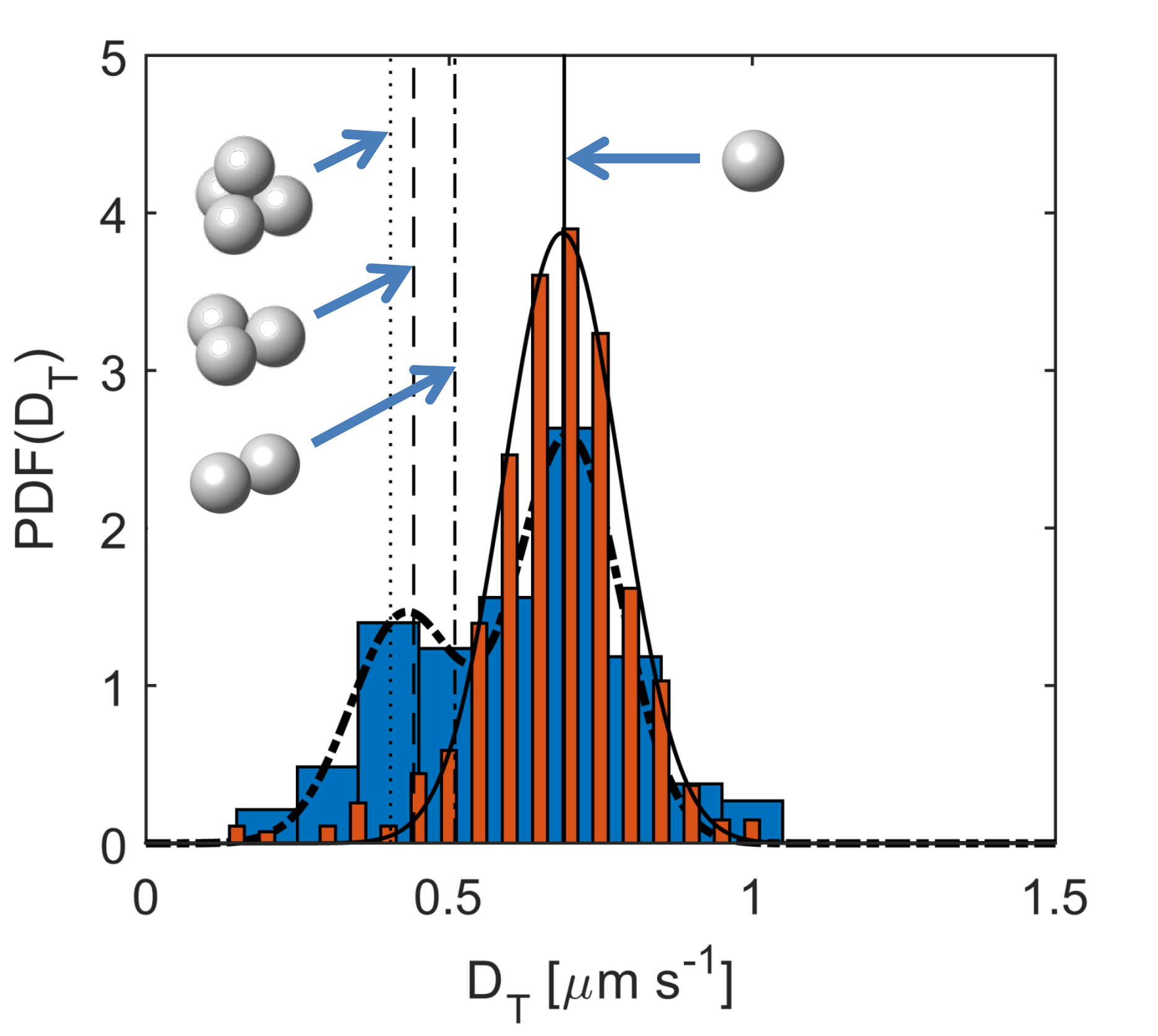}\caption{\textcolor{black}{Histograms of the distribution of translational
diffusion coefficient for Sample 1 (large blue bars) and for Sample
2 (thin orange bars) as determined from VPT. The continous vertical
line correspond to $\bar{D}_{T}=0.69$ $\mu m^{2}s^{-1}$, obtained
by fitting the histogram obtained from Sample 2 to a Gaussian function
(continous curve). The histogram for Sample 1 is well described as
the sum of two Gaussian functions }(heavy dashed-dotted curve)\textcolor{black}{{}
centered in} $D_{T}\simeq0.7$ $\mu m^{2}s^{-1}$ and $D_{T}\simeq0.4$
$\mu m^{2}s^{-1}$, respectively. This second value\textcolor{black}{{}
falls within the range of }the translational diffusion coefficients
expected for small clusters of monomers, assuming $\bar{D}_{T}$ as
the diffusion coefficient of a monomer. The vertical dashed-dotted,
dashed and dotted lines correspond to the expected values for dimers,
trimers and tetramers, respectively.\textcolor{black}{{} }\label{fig:AMPS-1-1}}
\end{figure}

\section{Conclusions}

In this work, we have shown that the recently introduced Dark Field
Differential Dynamic Microscopy \cite{Bayles:2016_darkDDM} is a simple
and powerful tool for the simultaneous determination of the roto-translational
dynamics of anisotropic microparticles in suspension. Our experiments
with bacterial suspensions showed that d-DDM is somehow complementary
to the recently proposed polarized-DDM \cite{Giavazzi:2016_jpcm},
in that the latter may fail with particles characterized by moderate
shape anisotropy and good refractive index-matching with the dispersion
medium.

We also showed that the peculiar nature of the dark field signal associated
with the rotational dynamics, makes d-DDM very effective in spotting
and quantifying the presence of anisotropic aggregates of isotropic
particles, even though whenever the particles are large enough, particle
tracking may provide a better tool for a detailed analysis of aggregated
samples in which clusters are present. On the other hand, dDDM analysis
is statistically more robust and does not require the intervention
of an experienced user for the fine tuning of the parameters involved
in the image processing procedure.

These promising results open to the possibility of studying with d-DDM
more complex and challenging systems. For example, it is know that,
for motile bacteria, the rotational dynamics is very different from
the purely Brownian one \cite{SARAGOSTI2012} and that a non-trivial
interplay exists between rotational and transational degrees of freedom
\cite{patteson2015}. d-DDM could allow the simple high-throughput,
characterization of this complex dynamics, possibly combination with
other quantitative microscopy methods, like for example standard phase
contrast DDM \cite{Martinez:2012ya} or the so-called dark field flicker
microscopy \cite{Martinez16122014}, that has been used for monitoring
the rapid periodic fluctuation associated with the beating of flagella. 

Another field of potential application is the optical characterization
of the mechanical properties of soft materials, the realm of microrheology
\cite{Cicuta:2007vn,Waigh:2016sf}. Passive microrheology, in particular,
exploits the thermally excited positional fluctuations of immersed
tracer particles to probe the viscoelastic moduli of the hosting fluid
\cite{Mason1995}. Very recently, DDM has been demonstrated to be
a reliable route to microrheology, enabling the accurate, tracking-free
determination of the mean squared displacement of probe particles
in a variety of imaging conditions \cite{Bayles2017,edera_2017}.
In view of the results presented in this work, we expect that d-DDM,
in combination with calibrated anisotropic tracers, could provide
the ideal ground to extend these ideas also to the rotational degree
of freedom, enabling the simultaneous execution of translational and
rotational microrheology experiments \cite{Cheng2003,Reyes2005}.
While this combination could appear redundant in the case of a perfectly
homogeneous fluid - where the two approaches are expected to give
equivalent results - it could quite valuable in the presence of sources
of non-ideality in the system (\emph{e.g.} inhomogeneity of the matrix
or specific tracer-fluid interactions altering the boundary conditions
at the surface of the particles). Since these effects, that can seriously
compromise the reliability microrheology results, are expected to
have a different impact on rotational and translational degrees of
freedom, the execution of a combined experiment could allow to spot
them effectively.

Moreover, since d-DDM is very sensitive in detecting the presence
of anisotropic particles, it can be exploited as a quality control
step during the preparation or the execution of experiments involving
allegedly spherical colloidal particles or for real-time monitoring
of aggregation processes and self-assembly. Another intersting application
of the method could be in the on-line moniting of water quality \cite{water_2016}.
In this case, the ability of d-DDM to spot the presence of particles
in solution and, by providing an estimate of their dimensions and
optical contrast, to discriminate between them, could make it an useful
screening tool for the automatic identification of particularly important
classes of contaminants, \emph{in primis} bacteria. 

\ack{}

\textcolor{black}{We thank Silvia Biffi, Filippo Saglimbeni and Giovanni
Tagliabue for help in the preparation of the samples and for early
contributions to the project. We also thank Roberto di Leonardo and
Claudio Maggi for insightful comments and discussions. RC, DP and
FG acknowledge funding by the Italian Ministry of Education and Research,
Futuro in Ricerca Project ANISOFT (RBFR125H0M) and by Fondazione CARIPLO-Regione
Lombardia Project Light for Life (2016-0998). MB acknowledges funding
from the European Union\textquoteright s Seventh Framework Programme
(FP7) for Research, Technological Development and Demonstration through
the NAPES project (grant agreement no. 604241). }

\appendix

\section{Roto-translational diffusion of spheroids and cylinders \ref{eq:normalisf}}

We report here analytic expression available in the literature for
the rotational and translational diffusion coefficient for a spheroidal
particle \cite{KOENIG} and for a cylinder \cite{TIRADO1984,PhysRevE.50.1232_lowen}: 

\subsubsection*{Spheroid}

The rotational and translational diffusion coefficients are given
by, respectively

\[
D_{R}=\frac{3k_{B}T}{32\pi\eta}\frac{\left(2a^{2}-b^{2}\right)S-2a}{a^{4}-b^{4}},
\]

\[
D_{T}=\frac{k_{B}T}{12\pi\eta}S.
\]

Here $S=(2/k)\ln\left(\frac{a+k}{b}\right))$, $k^{2}=a^{2}-b^{2}$,
$a$ and $b$ are the minor and the major semi-axis of the particle,
respectively, $k_{B}$is the Boltzmann constant, $T$ is the absolute
temperature and $\eta$ is the solvent viscosity.

\subsubsection*{Cylinder}

The rotational and translational diffusion coefficients are given
by, respectively

\[
D_{R}=\frac{3D_{0}}{\pi L^{2}}\left(\ln p-0.662+0.917/p-0.050/p^{2}\right),
\]

\[
D_{T}=\left(D^{\parallel}+2D^{\perp}\right)/3,
\]

where $D^{\perp}$and $D^{\parallel}$ are the translational diffusion
coefficients along the direction perpendicular and parallel to the
axis of the particle, respectively:

\[
D^{\perp}=\frac{D_{0}}{4\pi}\left(\ln p+0.839+0.185/p+0.233/p^{2}\right),
\]

\[
D^{\parallel}=\frac{D_{0}}{2\pi}\left(\ln p-0.207+0.980/p-0.133/p^{2}\right).
\]

In the above expression $D_{0}=k_{B}T/\eta L$, $p=L/\sigma$. $\sigma$
and $L$ are, respectively, the diameter and the length of the cylinder.
All other quantities are defined as in case of the spheroid.

\bibliographystyle{iopart-num}
\addcontentsline{toc}{section}{\refname}

\providecommand{\newblock}{}


\end{document}